\documentclass[aps,pre,reprint,amsmath,amssymb,amsfonts,superscriptaddress]{revtex4-2}

\usepackage{graphicx}
\usepackage{bm}

\usepackage[unicode=true,
 bookmarks=true,bookmarksnumbered=false,bookmarksopen=false,
 breaklinks=false,pdfborder={0 0 0},pdfborderstyle={},backref=false,colorlinks=true]
 {hyperref}
\hypersetup{pdfborderstyle={},pdfborderstyle={},pdfborderstyle={},pdfborderstyle={},pdfborderstyle={},pdfborderstyle={},linkcolor=blue,citecolor=blue}

   \usepackage{mathpazo}

    \newcommand\beq{\begin{equation}}
	\newcommand\eeq{\end{equation}}
\newcommand\beqa{\begin{eqnarray}}
	\newcommand\eeqa{\end{eqnarray}}
\newcommand{\nn}{\nonumber\\}
\def\bal#1\eal{\begin{align}#1\end{align}}

\newcommand{\bp}{\beta P}
\newcommand{\xior}{\xi_{\text{or}}}
\newcommand{\xipos}{\xi_{\text{pos}}}
\newcommand{\varphip}{\breve{\varphi}}
\newcommand{\varphim}{\bar{\varphi}}
\newcommand{\cont}{\text{cont}}
\newcommand{\tm}{\text{TM}}

\usepackage[dvipsnames]{xcolor}
\usepackage[normalem]{ulem}

\begin{document}

\title{Orientational ordering and correlations in a quasi-one-dimensional hard-dumbbell fluid}

\author{Ana M. Montero}
  \email{anamontero@unex.es}
  \affiliation{Departamento de F\'isica,
    Universidad de Extremadura, E-06006 Badajoz, Spain}
\author{P\'eter Gurin}
  \email{gurin.peter@mk.uni-pannon.hu}
  \affiliation{Physics Department, Centre for Natural Sciences,
    University of Pannonia, P.O. Box 158, Veszpr\'em H-8201, Hungary}
\author{Szabolcs Varga}
 \email{varga.szabolcs@mk.uni-pannon.hu}
 \affiliation{Physics Department, Centre for Natural Sciences,
   University of Pannonia, P.O. Box 158, Veszpr\'em H-8201, Hungary}
   \author{Andr\'es Santos}
  \email{andres@unex.es}
  \affiliation{Departamento de F\'isica,
    Universidad de Extremadura, E-06006 Badajoz, Spain}
  \affiliation{Instituto de Computaci\'on Cient\'ifica Avanzada (ICCAEx),
    Universidad de Extremadura, E-06006 Badajoz, Spain}

\date{\today}

\begin{abstract}
We study a quasi-one-dimensional fluid of hard dumbbells with continuous orientational degrees of freedom using an exact transfer-matrix formulation. The model allows for a complete analytical characterization of thermodynamic properties, orientational ordering, and correlation functions in terms of the spectral properties of an integral operator. We derive exact expressions for the equation of state, the orientational distribution function, and both partial and total radial distribution functions. Their asymptotic behavior is governed by the complex poles of the Laplace-transformed correlation functions, which determine the positional and orientational correlation lengths. At high densities, an extended intermediate regime with an approximate algebraic decay of the radial distribution function precedes the pole-dominated asymptotic exponential behavior. As density increases, the system exhibits a continuous crossover from a weakly ordered regime with a unimodal orientational distribution to a strongly constrained regime characterized by bimodal orientational ordering. This crossover is accompanied by a nonmonotonic behavior of the pressure relative to the Tonks gas and by a qualitative change in the decay of correlation functions from oscillatory to monotonic. In the high-pressure limit, we show that orientational and positional fluctuations contribute equally to the pressure, leading to a universal ratio of twice the Tonks pressure. The theoretical predictions are supported by numerical solutions of the discretized transfer operator and by scaling arguments that elucidate the high-pressure behavior of ordering and correlation lengths.
\end{abstract}

\maketitle

\section{Introduction}\label{intro}

Hard-body models with continuously degenerate close-packed structures display nontrivial behavior under extreme compression, driven by the competition between geometric constraints and entropy. A paradigmatic example is the one-dimensional system of hard needles, which has been studied in detail since the seminal work of Onsager~\cite{Onsager_AnnNYAcadSci_1949} and extensively revisited over the past decades~\cite{Kantor_2009,Kantor-Kardar_EPL_2009,Gurin_PRE_2011,Klamser-Sadhu-Dhar_PhysRevE_2022}.
In this model, the absence of a finite transverse size represents a strong idealization, which leads to unusual features such as a density that diverges with increasing pressure and orientational correlations that are of limited physical relevance. Consequently, hard-needle models have limited applicability to experimentally realizable systems.

The purpose of this work is to investigate a minimal hard-particle system that preserves the key geometric property of a continuously degenerate close packing, while avoiding the pathological aspects associated with vanishing thickness. To this end, we introduce a finite-width generalization of the hard-needle model by considering tangent hard dumbbells. The presence of a finite transverse size alters the packing constraints in a fundamental way, leading to qualitative changes in the equation of state, orientational ordering, and correlation lengths at high pressure.

\begin{figure}
\resizebox{0.48\textwidth}{!}{\includegraphics{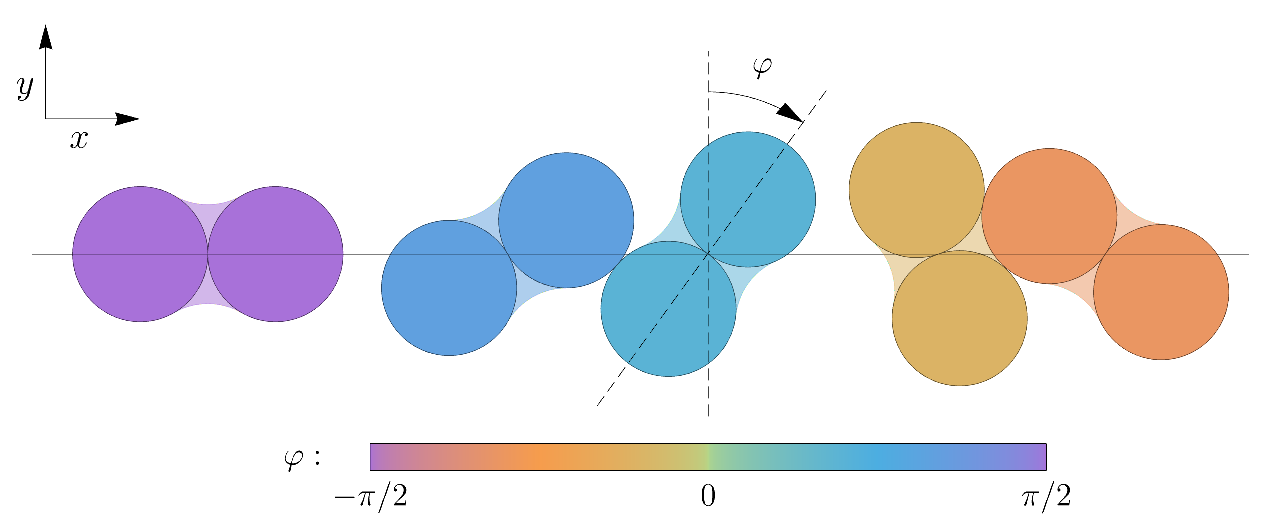}}
\caption{
Geometry of the q1D dumbbell system. The centers of mass are constrained to move along the $x$ axis, while each dumbbell can rotate in the $xy$ plane with orientation angle $\varphi$. Colors encode the dumbbell orientations. The light shading indicates the maximal admissible region that can connect the two disks of a dumbbell without affecting the contact distance between neighboring dumbbells, as illustrated in the rightmost pair.
\label{fig:dumbbells}}
\end{figure}

Specifically, we study a quasi-one-dimensional (q1D) system of hard dumbbells whose centers are constrained to lie along the $x$ axis, while they are free to rotate in the $xy$ plane (Fig.~\ref{fig:dumbbells}). This geometry constitutes a natural extension of the hard-needle problem: translational motion remains effectively one-dimensional, but each dumbbell carries a continuous orientational degree of freedom associated with its finite aspect ratio. As in the hard-needle case, the close-packed configurations are continuously degenerate; however, the finite size of the dumbbells introduces a nontrivial coupling between positional and orientational fluctuations.

From a geometric point of view, the contact distance between two adjacent dumbbells---defined as the minimal center-to-center separation along the channel at which the dumbbells touch without overlapping---does not require the two disks forming a dumbbell to be connected solely at a single point.
The two disks may instead be joined by a thicker neck of arbitrary shape, provided that this connecting region remains entirely within the circular arc traced by a third disk tangent to both constituent disks. The maximal admissible connecting region is indicated by the light shading in Fig.~\ref{fig:dumbbells}. Consequently, the analysis presented in this paper applies not only to ideal tangent dumbbells but also to a broader family of hard particles with similar nonconvex shapes.

This type of q1D systems is appealing for several reasons. They provide a controlled setting in which positional and orientational degrees of freedom are strongly intertwined, while still remaining amenable to exact or semi-analytical treatment; by contrast, the corresponding theoretical approaches become substantially more difficult in higher dimensions~\cite{Kulossa-Weidig-Wagner_PhysRevE_2023}. Moreover, they serve as minimal theoretical models for confined anisotropic particles in narrow channels, where translational motion is effectively one-dimensional but rotations are still allowed. Such systems are usually modeled using hard cylinders, spherocylinders, or ellipsoids~\cite{Wensink-Lowen..._EurPhysJ_2013,Luis-Kike-Yuri_JPhysCondensMatter_2014}; however, these shapes possess a nondegenerate, unique close-packed structure and therefore exhibit behavior near close packing that is qualitatively different from that of needles. Understanding how close-packing degeneracy manifests itself in q1D systems sheds light on the role of entropy in determining the equation of state, the structure of correlation functions, and the nature of ordering crossovers under strong confinement.

It is also important to note that a variety of synthesis techniques have been successfully employed to produce colloidal particles with dumbbell-like shapes~\cite{Johnson..._Langmuir_2005,Mock..._Langmuir_2006,Lee..._JMaterChem_2008}. The bulk phase behavior of hard dumbbells is well established from simulation studies~\cite{Vega-Paras-Monson_JChemPhys.96_1992,Vega-Paras-Monson_JChemPhys.97_1992,Vega-Monson_JChemPhys_1997,Marechal-Dijkstra_PhysRevE_2008}, and their quasi-two-dimensional behavior has also been investigated~\cite{Gerbode-Lee-Liddell-Cohen_PhysRevLett_2008,Gerbode-...-Escobedo_PhysRevLett_2010}. By contrast, in q1D confinement hard dumbbells have so far been studied only with discretized orientations~\cite{Montero-Santos-Gurin-Varga_JCP_2023}. Owing to their nonconvex shape, dumbbells have also served as a valuable test case for extended deconvolution fundamental measure theory~\cite{Marechal-Goetzke-Hartel-Loeven_JChemPhys_2011}, and more recently they have attracted interest as active particles~\cite{Venkatareddy-Lin-ShiangTai-Maiti_PhysRevE_2023}.

As the density increases, the system develops pronounced orientational ordering associated with the geometry of the dumbbells and the degeneracy of the close-packed configurations. The resulting ordering corresponds to a continuous reorganization of the orientational distribution from a single preferred orientation to two symmetric ones. Although this behavior bears some resemblance to a nematic--tetratic transition, in the present q1D system with nearest-neighbor interactions it should be interpreted as a smooth structural crossover rather than as a genuine thermodynamic phase transition.

This paper is organized as follows. In Sec.~\ref{sec2}, we define the model and detail the geometrical constraints imposed by q1D confinement. In Sec.~\ref{sec3}, we develop the theoretical framework used to obtain exact expressions for thermodynamic and structural quantities. The resulting equation of state, orientational distribution function, and correlation functions are presented and discussed in Sec.~\ref{sec4}. Finally, Sec.~\ref{sec5} summarizes the main findings and outlines possible extensions of the present work.

\section{Dumbbell model}\label{sec2}

In our model (see Fig.~\ref{fig:dumbbells}), each particle is a hard dumbbell composed of two hard disks of diameter $a$, lying in the $xy$ plane and rigidly connected at a common point that defines the particle's center of symmetry. This center is constrained to move along the $x$ axis, rendering the system q1D. Each dumbbell can freely rotate in the $xy$ plane, its orientation being specified by the angle $\varphi$ between the dumbbell axis and the $y$ axis. By symmetry, the angular domain is restricted to $-\pi/2 \le \varphi \le \pi/2$.

Interactions are limited to nearest neighbors. The hard-body constraint is fully characterized by the contact distance $\sigma(\varphi,\varphi')$ between the centers of two adjacent dumbbells, which depends on their respective orientations $\varphi$ and $\varphi'$.

Suppose a dumbbell is located at position $x$ with orientation $\varphi$. The coordinates of the centers of its upper ($u$) and lower ($l$) disks are
\beq
x_{\alpha}=x+s_{\alpha} \frac{a}{2}\sin \varphi,\quad y_{\alpha}=s_{\alpha}\frac{a}{2}\cos \varphi,\quad \alpha=u,l,
\eeq
where $s_u=+1$ and $s_l=-1$.
Consider now a second dumbbell at position $x'>x$ with orientation $\varphi'$. The two dumbbells may come into contact through their upper disks ($uu$), their lower disks ($ll$), or through one upper and one lower disk ($ul$ or $lu$). The contact condition is
\beq
a^2=(x_{\alpha'}-x_\alpha)^2+(y_{\alpha'}-y_\alpha)^2,
\eeq
which implies that the center-to-center separation at contact is $x'-x = a\, d_{\alpha\alpha'}(\varphi,\varphi')$, where the dimensionless function $d_{\alpha\alpha'}(\varphi,\varphi')$ is given by
\bal
d_{\alpha\alpha'}(\varphi,\varphi')=&\sqrt{1-\frac{1}{4}(s_\alpha\cos\varphi-s_{\alpha'}\cos\varphi')^2}\nn
&+\frac{1}{2}(s_\alpha\sin\varphi-s_{\alpha'}\sin\varphi'),\quad \alpha,\alpha'=u,l.
\eal
Taking into account all four possible contact configurations ($uu$, $ll$, $ul$, and $lu$), the contact distance between the centers of two neighboring dumbbells is
  \begin{equation}
    \sigma(\varphi,\varphi')
  = a \max_{\alpha,\alpha'} \left[ d_{\alpha\alpha'}(\varphi, \varphi') \right].
  \label{eq:sigma_dumbbell}
  \end{equation}
Note that the contact distance $\sigma(\varphi,\varphi')$ satisfies the bisymmetry property
\beq
\label{bisymm}
\sigma(\varphi,\varphi')=\sigma(\varphi',\varphi)=\sigma(-\varphi,-\varphi'),
\eeq
which follows directly from the symmetry of the dumbbell geometry under particle exchange and under a simultaneous inversion of both orientations.

\begin{figure}
    \resizebox{0.48\textwidth}{!}{\includegraphics{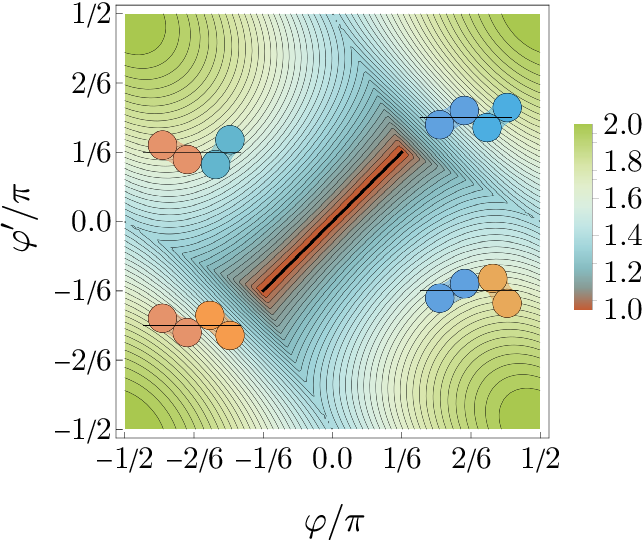}}
       \caption{
       Contour plot of the reduced contact distance $\sigma(\varphi,\varphi')/a$ for two dumbbells. The black line marks the locus $-\pi/6 \le \varphi=\varphi' \le \pi/6$, where the contact distance attains its minimum value $\sigma=a$. The behavior of $\sigma$ in the immediate vicinity of this line governs the high-pressure properties of the system. The maximum value, $\sigma=2a$, occurs at the corners $\varphi,\varphi'=\pm\pi/2$. The four dumbbell pairs illustrate representative configurations from the four regions of the diagram. From the top-left region and proceeding clockwise, the corresponding coordinates are $(\varphi,\varphi')=(-\pi/3,\pi/6)$, $(\pi/3,\pi/4)$, $(\pi/3,-\pi/6)$, and $(-\pi/3,-\pi/4)$.
      \label{fig:contact_distance_dumbbell}}
        \end{figure}

\begin{figure}
        \resizebox{0.4\textwidth}{!}{\includegraphics{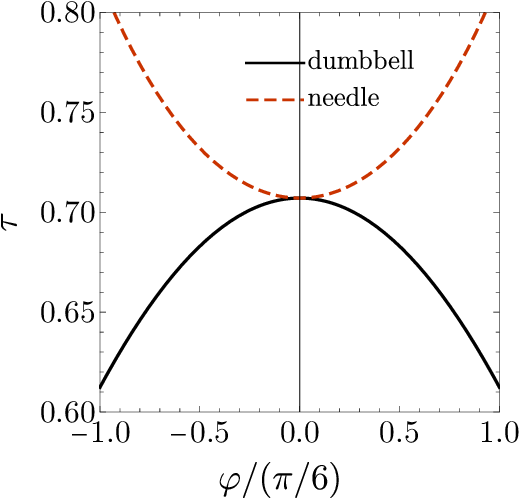}}
      \caption{
      The solid line shows the slope of the contact distance along the valley,
$\tau \equiv \lim_{\varphi'\rightarrow\varphi}|\bm{\nabla} \sigma(\varphi,\varphi')/a|=(\cos\varphi)/\sqrt{2}$,
for $-\pi/6<\varphi=\varphi'<\pi/6$. Since $\bm{\nabla}\sigma$ along the valley is orthogonal to the segment $-\pi/6<\varphi=\varphi'<\pi/6$, $\tau$ coincides with the directional derivative of $\sigma$ along that direction. The dashed line shows the corresponding quantity for needles.
      \label{fig:slope_of_contact_distance}}
  \end{figure}

Figure~\ref{fig:contact_distance_dumbbell} illustrates the dependence of the contact distance $\sigma(\varphi,\varphi')$ on the orientations $\varphi$ and $\varphi'$. The $(\varphi,\varphi')$ plane is partitioned into four regions, delimited by the curves $\sin\varphi=(\cos\varphi')/\sqrt{3}$ for $0\le\varphi\le\pi/6$, $\sin\varphi'=(\cos\varphi)/\sqrt{3}$ for $\pi/6\le\varphi\le\pi/2$, $\sin\varphi=-(\cos\varphi')/\sqrt{3}$ for $-\pi/6\le\varphi\le 0$, and $\sin\varphi'=-(\cos\varphi)/\sqrt{3}$ for $-\pi/2\le\varphi\le -\pi/6$, together with the diagonal segment $-\pi/6\le\varphi=\varphi'\le\pi/6$. Starting from the top-left region and proceeding clockwise, the dominant contributions to the contact distance arise from $d_{ll}$, $d_{ul}$, $d_{uu}$, and $d_{lu}$, respectively.

We observe that $\sigma(\varphi,\varphi')$ exhibits a continuously degenerate minimum along the line $\varphi=\varphi'$ for $-\pi/6\le\varphi\le\pi/6$. In this respect, dumbbells behave similarly to needles, while incorporating a finite-thickness geometry.

However, important differences remain between the two particle shapes. First, for needles the contact distance attains its minimum for any orientation satisfying $\varphi=\varphi'$, whereas for dumbbells this condition is restricted to the finite interval $-\pi/6\le\varphi=\varphi'\le\pi/6$. Second, and more importantly for the high-pressure properties of the system, the two models differ in the behavior of the contact distance $\sigma$ in the immediate vicinity of the minimum line. This behavior is quantified by the magnitude of the gradient
\beq
\tau(\varphi)\equiv \lim_{\varphi'\to\varphi}\left|\bm{\nabla} \sigma(\varphi,\varphi')/a\right|
=\frac{\cos\varphi}{\sqrt{2}},
\eeq
evaluated along the ``valley'' $-\pi/6<\varphi=\varphi'<\pi/6$. Note that, although $\sigma(\varphi,\varphi')$ is not differentiable at points on the line $-\pi/6\le\varphi=\varphi'\le\pi/6$, the quantity $\tau$ is nevertheless well defined everywhere along the valley except at the endpoints $\varphi=\varphi'=\pm\pi/6$.

The slope $\tau(\varphi)$ is plotted in Fig.~\ref{fig:slope_of_contact_distance}, alongside the corresponding quantity for needles, $\tau_{\rm needle}=(\sec\varphi)/\sqrt{2}$. For needles, the slope has a single minimum, $\tau_{\rm needle}=1/\sqrt{2}$ at $\varphi=\varphi'=0$, and increases monotonically away from this point. In contrast, the dumbbell system exhibits two symmetric minima, $\tau=\sqrt{3/2}/2$, when $\varphi=\varphi'$ and $|\varphi|\to(\pi/6)^-$.
This twofold degeneracy produces high-pressure behavior markedly different from that of needles. Near $\varphi=\varphi'=\pm\pi/6$, the reduced slope of $\sigma(\varphi,\varphi')$ allows greater orientational fluctuations. Consequently, the orientational distribution function (ODF), $f(\varphi)$, is expected to develop two peaks, one just above $-\pi/6$ and one just below $\pi/6$, as the density exceeds a certain threshold. This behavior will be further analyzed in Sec.~\ref{sec4}.

\section{Theoretical framework}
\label{sec3}

\subsection{Transfer-matrix method}
A key feature of the system is its q1D nature, in which each dumbbell interacts only with its immediate left and right neighbors. Under this condition, the total potential energy of a system of $N$ dumbbells can be written as
\beq
\label{5}
U_N(\{x_i,\varphi_i\})=\sum_{i=1}^N u(r_{i};\varphi_i,\varphi_{i+1}),
\eeq
where $r_{i}\equiv x_{i+1}-x_i$, and the pair potential is given by
\beq
\label{6}
u(r;\varphi,\varphi')=\begin{cases}
\infty,& r<\sigma(\varphi,\varphi'),\\
0,& r>\sigma(\varphi,\varphi').
\end{cases}
\eeq
In this expression, we have assumed periodic boundary conditions, so that $x_{N+1}=x_1+L$ and $\varphi_{N+1}=\varphi_1$, where $L$ is the system length. The Boltzmann factor then factorizes as
\beq
    e^{-\beta U_N(\{x_i,\varphi_i\})} =
              \prod_{i=1}^N e^{-\beta u(r_{i};\varphi_i,\varphi_{i+1})} ,
   \label{micro_Boltzmann}
  \eeq
where $\beta=1/k_BT$ is the inverse temperature.

In the absence of external potentials, the $N$-body configurational probability density reads
\bal
\label{micro_probability}
p_N(\{x_i,\varphi_i\})=&\frac{ e^{-\beta U_N(\{x_i,\varphi_i\})-\bp L}}{Z_N}\nn
=&\prod_{i=1}^N p_1(r_i;\varphi_i,\varphi_{i+1}),
\eal
where $P$ is the pressure and
\beq
\label{p1}
p_1(r;\varphi,\varphi')=\frac{e^{-\beta u(r;\varphi,\varphi')-\bp r}}{Z_N^{1/N}}
\eeq
is the one-bond contribution. The
configurational part of the isothermal-isobaric partition function becomes
\bal
\label{Z}
Z_N=&\int_0^\infty dL\, e^{-\beta PL}\int_0^L d x_1\int d\varphi_1 \int_{x_1}^L d x_2\int d\varphi_2 \nn
&\times \cdots\int_{x_{N-1}}^L dx_N\int d\varphi_N\, e^{-\beta U_N(\{x_i,\varphi_i\})}\nn
=&\int d\varphi_1\int d\varphi_2\cdots\int d\varphi_N\, \prod_{i=1}^N K(\varphi_i,\varphi_{i+1})\nn
=&\text{Tr}\widehat{K}^N.
\eal
Here,
  \beq
    K(\varphi,\varphi')
    =
    \int_0^\infty d r\, e^{-\beta u(r;\varphi,\varphi')-\beta P r},
    \label{K}
  \eeq
and $\widehat{K}$ is the integral operator with kernel $K(\varphi,\varphi')$, defined by
\beq
\widehat{K}\Psi(\varphi)=\widehat{K}|\Psi\rangle=\int d\varphi' \, K(\varphi,\varphi')\Psi(\varphi').
\eeq
In the derivation of Eq.~\eqref{Z} we have used the standard relation $\int_0^L d x_1\cdots \int_{x_{N-1}}^L dx_N=\int_0^\infty dr_1\cdots \int_0^\infty dr_N\, \delta(L-\sum_i r_i) $,
which transforms the ordered integrals over particle positions into integrals over nearest-neighbor separations.

Equation~\eqref{Z} provides the foundation of the transfer-operator method, where $\widehat{K}$ is referred to as the transfer operator~\cite{Lebowitz-Percus-Talbot_JStatPhys_1987,Kofke-Post_JCP_1993,Varga-Ballo-Gurin_JStatMechTheorExp_2011}. A key advantage of this approach is that the trace of $\widehat{K}^N$ can be evaluated in any convenient basis. In particular, consider the basis ${\psi_k(\varphi)}$ formed by the eigenfunctions of the transfer operator,
\beq
\widehat{K}|\psi_k\rangle=\lambda_k|\psi_k\rangle\Leftrightarrow \int d\varphi' \, K(\varphi,\varphi')\psi_k(\varphi')=\lambda_k\psi_k(\varphi),
\eeq
where $\{\lambda_k\}$ are the corresponding eigenvalues.
The bisymmetry property of Eq.~\eqref{bisymm} implies that $\widehat{K}$ itself is bisymmetric, which in turn ensures that its eigenfunctions have well-defined parity:
\beq
\psi_k(\varphi) = (-1)^k \psi_k(-\varphi).
\eeq

If the eigenfunctions are normalized, the corresponding eigenvalues can be written as
\beq
\label{12}
\lambda_k=\langle \psi_k|\widehat{K}|\psi_k\rangle,
\eeq
where the scalar product is defined as
\beq
\langle \Phi |\Psi\rangle=\int d\varphi \, \Phi(\varphi)\Psi(\varphi).
\eeq
Equation~\eqref{Z} then implies $Z_N=\sum_{k=0}^\infty \lambda_k^N$, and in the thermodynamic limit, $Z_N \to \lambda_0^N$, where $\lambda_0$ is the largest eigenvalue. Therefore, determining $\lambda_0$ is sufficient to characterize the system's thermodynamics. Moreover, under rather general conditions, the Perron--Frobenius--Jentzsch theorem guarantees the existence of a unique dominant eigenvalue $\lambda_0$~\cite{Lebowitz-Percus-Talbot_JStatPhys_1987}.

Thermodynamic quantities and the equation of state follow directly from the configurational integral. The Gibbs free energy per particle is
$G/N=-(k_B T/N) \ln Z_N\to -k_BT\ln\lambda_0$ in the thermodynamic limit. Hence the number density satisfies $\rho^{-1}=\partial (G/N)/\partial P=-\partial_{\bp} \ln\lambda_0$. From Eq.~\eqref{12}, we have
\beq
\frac{\partial \lambda_0}{\partial\bp}=\left\langle\psi_0\left|\frac{\partial\widehat{K}}{\partial\bp}\right|\psi_0\right\rangle +2\left\langle\frac{\partial\psi_0}{\partial\bp}\left|\widehat{K}\right|\psi_0\right\rangle.
\eeq
The second term vanishes because $2\langle\partial_{\bp}\psi_0|\widehat{K}|\psi_0\rangle=\lambda_0\partial_{\bp}\langle\psi_0|\psi_0\rangle=0$. Thus the equation of state can be written as
\beq
\label{17}
\frac{1}{\rho}=\frac{\langle\psi_0|\widehat{K}'|\psi_0\rangle}{\langle\psi_0|\widehat{K}|\psi_0\rangle},
\eeq
where $\widehat{K}'\equiv -\partial_{\bp}\widehat{K}$ is the operator with kernel
\bal
K'(\varphi,\varphi')=&-\frac{\partial K(\varphi,\varphi')}{\partial\bp}\nn
=&\int_0^\infty d r\, r e^{-\beta u(r;\varphi,\varphi')-\beta P r}.
\eal

The marginal probability density for the orientation of a single particle, $f(\varphi_1)$, is obtained by integrating the $N$-body probability distribution $p_N(\{x_i,\varphi_i\})$ in Eq.~\eqref{micro_probability} over all positions and over the orientations of all particles except particle $1$. This yields
 \bal
f(\varphi_1)=&\frac{\int d\varphi_2\int d\varphi_3\cdots\int d\varphi_N\,\prod_{i=1}^N K(\varphi_i,\varphi_{i+1})}{\lambda_0^N}\nn
=&\frac{K^N(\varphi_1,\varphi_1)}{\lambda_0^N},
\eal
where $K^N(\varphi,\varphi')$ is the kernel of the operator $\widehat{K}^N$.

Using the spectral representation $K(\varphi,\varphi')=\sum_{k=0}^\infty\lambda_k\psi_k(\varphi)\psi_k(\varphi')$, one has  $K^N(\varphi,\varphi')=\sum_{k=0}^\infty\lambda_k^N\psi_k(\varphi)\psi_k(\varphi')\to \lambda_0^N\psi_0(\varphi)\psi_0(\varphi')$ in the thermodynamic limit. Thus, the ODF becomes
\beq
\label{ODF}
f(\varphi)=\psi_0^2(\varphi).
\eeq
This representation allows the average of any physical quantity that depends only on a single-particle orientation, $A(\varphi)$, to be expressed as
  \begin{equation}
    \langle A \rangle
    =  \langle\psi_0 | A | \psi_0\rangle=\int d\varphi\,A(\varphi)\psi_0^2(\varphi).
  \end{equation}

A useful measure of orientational ordering is the nematic order parameter, defined as
\beq
\label{order_param_S2}
S_2=\langle e^{2\imath\varphi}\rangle=\langle \cos(2\varphi)\rangle,
\eeq
where in the second equality we have used the symmetry property $\psi_0(\varphi) = \psi_0(-\varphi)$, which implies $\langle \sin(2\varphi) \rangle = 0$.
In the present two-dimensional geometry the nematic director $\mathbf{n}=(\sin\varphi_{\mathbf n},\cos\varphi_{\mathbf n})$, with $\varphi_{\mathbf n}=\arg\langle e^{2\imath\varphi}\rangle/2$, is always perpendicular to the $x$ axis since $\varphi_{\mathbf n}=0$ at any density. However, as will be shown in Sec.~\ref{sec4.B}, in the high-density regime the ODF becomes strongly bimodal with peaks near $\varphi=\pm\pi/6$, so that practically no dumbbells are oriented close to $\varphi_{\mathbf n}=0$. In this sense, the nematic director becomes a  misleading descriptor of the orientational order.

To characterize this bimodal ordering we therefore introduce an additional ``hexatic'' order parameter,
\beq
\label{order_param_S6}
S_6=-\langle e^{6\imath\varphi}\rangle=-\langle \cos(6\varphi)\rangle.
\eeq
This quantity is analogous to the bond-order parameter used to characterize the hexatic phase in two-dimensional systems, but here it measures the orientational order of the dumbbells rather than that of their bonds. The minus sign in Eq.~\eqref{order_param_S6} is introduced so that $S_6>0$ in the high-density limit.

The average of a quantity that depends on the orientations of two particles---such as the orientation correlation function---can be obtained by inserting the spectral expansion of the transfer operator into the corresponding two-point average. In this case, however, the dominant eigenvalue and eigenfunction alone are not sufficient. One finds
  \begin{align}
    G_A(n)
  \equiv& \langle A(\varphi_i) A(\varphi_{i+n}) \rangle
      - \langle A(\varphi_i) \rangle \langle A(\varphi_{i+n}) \rangle
      \nonumber \\
    =& \sum_{k=1}^\infty \left( \frac{\lambda_k}{\lambda_0} \right)^n
       \langle\psi_0 | A | \psi_k \rangle \langle\psi_k | A | \psi_0\rangle.
                \end{align}
For large separations $n$, the term with the second-largest eigenvalue $\lambda_1$ dominates the sum, so that $G_A(n)\sim e^{-n/\xior^\tm}$, where  the transfer-matrix orientational correlation length is
\beq
\label{25}
\xior^\tm=\frac{1}{\ln|\lambda_0/\lambda_1|}.
\eeq

\subsection{Radial distribution function}
\label{sec3B}
Despite the advantages of the transfer-matrix method, once the positional coordinates $\{x_i\}$ have been integrated out, the method is not ideally suited for computing spatially dependent quantities such as the radial distribution function (RDF)  $g(r;\varphi,\varphi')$. Nevertheless, as in other q1D systems~\cite{MS23b,MS24,MS24b,MS25,ThesisAna}, one can exploit an exact mapping of the dumbbell system onto a polydisperse 1D mixture of hard rods, whose structural properties are exactly known~\cite{Salsburg-Zwanzig-Kirkwood_JCP_1953,Santos_LectureNotesInPhys_2016}.

The starting point is the first-neighbor probability distribution $p^{(1)}(r;\varphi,\varphi')$, which satisfies the normalization condition $\int_0^\infty dr\int d\varphi'\, p^{(1)}(r;\varphi,\varphi')=1$ for any $\varphi$. This function is not exactly given by the one-bond distribution in Eq.~\eqref{p1}, because $Z_N^{1/N}=\lambda_0=\langle\psi_0|\widehat{K}|\psi_0\rangle$, and thus $p_1(r;\varphi,\varphi')$ is not  normalized to unity. The proper normalization is achieved by defining
\bal
\label{21}
p^{(1)}(r;\varphi,\varphi')=&\frac{\psi_0(\varphi')}{\psi_0(\varphi)}p_1(r;\varphi,\varphi')\nn
=&\frac{\psi_0(\varphi')}{\lambda_0\psi_0(\varphi)}e^{-\beta u(r;\varphi,\varphi')-\bp r}.
\eal

The $n$th-neighbor probability distribution, $p^{(n)}(r;\varphi,\varphi')$, satisfies the convolution relation
\bal
\label{22}
p^{(n)}(r;\varphi,\varphi')=&\int_0^r dr'\,\int d\varphi''\,p^{(n-1)}(r';\varphi,\varphi'')\nn
&\times p^{(1)}(r-r';\varphi'',\varphi'),
\eal
and the pair correlation function follows from
\beq
\label{23}
\rho\psi_0^2(\varphi')g(r;\varphi,\varphi')=\sum_{n=1}^\infty p^{(n)}(r;\varphi,\varphi').
\eeq

In Laplace space, Eqs.~\eqref{21}--\eqref{23} become
\begin{subequations}
\beq
\label{21L}
\widetilde{P}^{(1)}(s;\varphi,\varphi')=\frac{\psi_0(\varphi')}{\lambda_0\psi_0(\varphi)}\widetilde{\Omega}(s+\bp;\varphi,\varphi'),
\eeq
\beq
\label{22L}
\widetilde{P}^{(n)}(s;\varphi,\varphi')=\int d\varphi''\,\widetilde{P}^{(n-1)}(s;\varphi,\varphi'') \widetilde{P}^{(1)}(s;\varphi'',\varphi'),
\eeq
\beq
\label{23L}
\rho\psi_0^2(\varphi')\widetilde{G}(s;\varphi,\varphi')=\sum_{n=1}^\infty \widetilde{P}^{(n)}(s;\varphi,\varphi').
\eeq
\end{subequations}
Here,
\beq
\widetilde{\Omega}(s;\varphi,\varphi')=\int_0^\infty dr\, e^{- sr}e^{-\beta u(r;\varphi,\varphi')}
\eeq
is the Laplace transform of the Boltzmann factor $e^{-\beta u(r;\varphi,\varphi')}$. Note that the kernel $K(\varphi,\varphi')$ of the transfer operator  coincides with $\widetilde{\Omega}(s=\bp;\varphi,\varphi')$.

Inserting Eq.~\eqref{22L} into Eq.~\eqref{23L}, we can write
\bal
\label{26}
\widetilde{G}(s;\varphi,\varphi')=&\int d\varphi'' \,\frac{\sum_{n=1}^\infty \widetilde{P}^{(n)}(s;\varphi,\varphi'')}{\rho\psi_0^2(\varphi')} \widetilde{P}^{(1)}(s;\varphi'',\varphi')\nn
&+\frac{\widetilde{P}^{(1)}(s;\varphi,\varphi')}{\rho\psi_0^2(\varphi')}\nn
=&\int d\varphi''\,\widetilde{G}(s;\varphi,\varphi'')
 \frac{\psi_0^2(\varphi'')}{\psi_0^2(\varphi')}
\widetilde{P}^{(1)}(s;\varphi'',\varphi')\nn
&+\frac{\widetilde{P}^{(1)}(s;\varphi,\varphi')}{\rho\psi_0^2(\varphi')}.
\eal
Finally, using Eq.~\eqref{21L}, this leads to the integral equation
\bal
\label{32}
\lambda_0\widetilde{\Gamma}(s;\varphi,\varphi')=
&\int d\varphi''\, \widetilde{\Gamma}(s;\varphi,\varphi'')\widetilde{\Omega}(s+\bp;\varphi'',\varphi')\nn
&+{\widetilde{\Omega}(s+\bp;\varphi,\varphi')},
\eal
where
\beq
\label{IE}
\widetilde{\Gamma}(s;\varphi,\varphi')\equiv \rho \psi_0(\varphi)\psi_0(\varphi')\widetilde{G}(s;\varphi,\varphi').
\eeq

In the language of operators, the formal solution of Eq.~\eqref{32} reads
\beq
\label{34}
\widehat{\Gamma}(s)=\left[\lambda_0\widehat{I}-\widehat{\Omega}(s+\bp)\right]^{-1}\cdot \widehat{\Omega}(s+\bp),
\eeq
where $\widehat{I}$ is the identity operator with kernel $\delta(\varphi-\varphi')$.

Let us denote by $\lambda_k(s)$ and $|\psi_k(s)\rangle$ the eigenvalues and eigenvectors, respectively, of the operator $\widehat{\Omega}(s)$. These represent the generalization  of the eigenvalues and eigenvectors of $\widehat{K}$ to the case $s\neq\bp$.
The spectral decomposition of $\widehat{\Gamma}(s)$  then reads
\beq
\widehat{\Gamma}(s)=\sum_{k=0}^\infty\frac{\lambda_k(s+\bp)}{\lambda_0-\lambda_k(s+\bp)}|\psi_k(s+\bp)\rangle\langle\psi_k(s+\bp)|.
\eeq
Thus, the nonzero poles \footnote{For simplicity, we will use the term ``pole'' to refer collectively to both real values and complex-conjugate pairs.} $\{s_{n}^{(k)}\}$ of $\widehat{\Gamma}(s)$ are the solutions of
\beq
\label{43.0}
\lambda_k(s_{n}^{(k)}+\bp)=\lambda_0,\quad k=0,1,2,\ldots,\quad n=1,2,3,\ldots.
\eeq
This allows the poles to be sorted into classes, each associated with a different eigenvalue $\lambda_k(s+\beta P)$. All poles have negative real parts, and within each class we order them according to $0>\text{Re}[s_1^{(k)}]\geq \text{Re}[s_2^{(k)}]\geq \text{Re}[s_3^{(k)}]\geq\cdots$.

The residues of $\widehat{\Gamma}(s)$ are
\bal
\widehat{R}_{n}^{(k)}=&\lim_{s\to s_{n}^{(k)}}(s-s_{n}^{(k)})\widehat{\Gamma}(s)\nn
=&-\frac{\lambda_0}{\lambda_k'(s_{n}^{(k)}+\bp)}|\psi_k(s_{n}^{(k)}+\bp)\rangle\langle\psi_k(s_{n}^{(k)}+\bp)|,
\eal
where $\lambda_k'(s)\equiv \partial\lambda_k(s)/\partial s$.
By application of the residue theorem, the RDF can be expressed as
\beq
\label{38}
g(r;\varphi,\varphi')=1+\sum_{k=0}^\infty\sum_{n=1}^\infty\frac{R_n^{(k)}(\varphi,\varphi')}{\rho\psi_0(\varphi)\psi_0(\varphi')}e^{s_n^{(k)}r}.
\eeq
The bisymmetry of $\widetilde{\Gamma}(s;\varphi,\varphi')$ further implies the parity property
\beq
\label{parity}
R_n^{(k)}(\varphi,-\varphi')=R_n^{(k)}(-\varphi,\varphi')=(-1)^k R_n^{(k)}(\varphi,\varphi').
\eeq

From the  partial RDFs $g(r;\varphi,\varphi')$, the total RDF  is obtained as
\beq
g(r)=\int d\varphi\int d\varphi'\,f(\varphi)f(\varphi')g(r;\varphi,\varphi').
\eeq
The parity property in Eq.~\eqref{parity} implies that only poles associated with even eigenvectors contribute to $g(r)$. Therefore, we can write
\bal
\label{41}
g(r)=&1+\frac{1}{\rho}\sum_{k=\text{even}}\sum_{n=1}^\infty\langle \psi_0|\widehat{R}_{n}^{(k)}|\psi_0\rangle e^{s_n^{(k)}r}\nn
=&1-\frac{\lambda_0}{\rho}\sum_{k=\text{even}}\sum_{n=1}^\infty\frac{|\langle \psi_0|\psi_k(s_n^{(k)}+\bp)\rangle|^2}{\lambda_k'(s_{n}^{(k)}+\bp)}e^{s_n^{(k)}r}.
\eal

Given a two-particle function $A(\varphi,\varphi')$, its average over all pairs of particles separated by a distance $r$ is
\beq
\langle A(\varphi,\varphi')\rangle_r=\frac{\int d\varphi\int d\varphi'\,f(\varphi)f(\varphi')g(r;\varphi,\varphi') A(\varphi,\varphi')}{g(r)}.
\eeq
Since the minimum separation between two particles is $r=1$, the average $\langle A(\varphi,\varphi')\rangle_r$ is ill-defined if $r<1$. In such cases, we adopt the convention  $\langle A(\varphi,\varphi')\rangle_{r<1}=0$.

We now define the distance-dependent orientational correlation function  as
\beq
G(r)=\langle \cos[2(\varphi-\varphi')]\rangle_r-S_2^2.
\eeq
In contrast to the total RDF $g(r)$, all poles (even and odd) contribute to $G(r)$.

Except for Eq.~\eqref{6}, all the results presented in this Section are valid for any q1D system in which the particles possess a continuous internal degree of freedom represented by the variable $\varphi$ and the kernel $\widetilde{\Omega}(s;\varphi,\varphi')$ is bisymmetric. For the specific case of our dumbbell system, where the pair potential is given by Eq.~\eqref{6}, one has
\begin{subequations}
\beq
\widetilde{\Omega}(s;\varphi,\varphi')=\frac{e^{-\sigma(\varphi,\varphi')s}}{s},
\eeq
\beq
K(\varphi,\varphi')=\frac{e^{-\sigma(\varphi,\varphi')\bp}}{\bp},
\eeq
\beq
K'(\varphi,\varphi')=K(\varphi,\varphi')\left[\sigma(\varphi,\varphi')+\frac{1}{\bp}\right].
\eeq
\end{subequations}
Accordingly, the equation of state, Eq.~\eqref{17}, can be written in the compact form
\beq
\label{EOS}
\frac{1}{\rho}=\frac{1}{\bp}+\frac{\langle\psi_0|\widehat{K}\sigma|\psi_0\rangle}{\langle\psi_0|\widehat{K}|\psi_0\rangle}.
\eeq
Here, the first term, $1/\beta P$, represents the ideal-gas-like contribution, while the second term accounts for the finite size of the dumbbells through the transfer-operator average of the contact distance.
Note that
\beq
\lim_{\bp\to 0}\frac{\langle\psi_0|\widehat{K}\sigma|\psi_0\rangle}{\langle\psi_0|\widehat{K}|\psi_0\rangle}=\langle\sigma\rangle_{\text{iso}}\simeq 1.5345 a,
\eeq
where $\langle\sigma\rangle_{\text{iso}}$ is the isotropically averaged contact distance. Thus, in the low-pressure limit, the equation of state reduces to
\beq
\bp=\rho\left[1+\rho\langle\sigma\rangle_{\text{iso}}+\mathcal{O}(\rho^2)\right].
\eeq

\section{Results}
\label{sec4}

We now present our results for the dumbbell model introduced in Sec.~\ref{sec2}. Since the orientation angle $\varphi$ is a continuous variable, all calculations are performed by discretizing the interval $-{\pi}/{2} \le \varphi \le {\pi}/{2}$ into a large number $M$ of points (typically $M = 500$--$1000$) and subsequently extrapolating to the continuum limit, $M \to \infty$. Throughout, we adopt $a = 1$ as the unit of length and $k_B T = 1/\beta = 1$ as the unit of energy.

\begin{figure}
  \resizebox{0.4\textwidth}{!}{\includegraphics{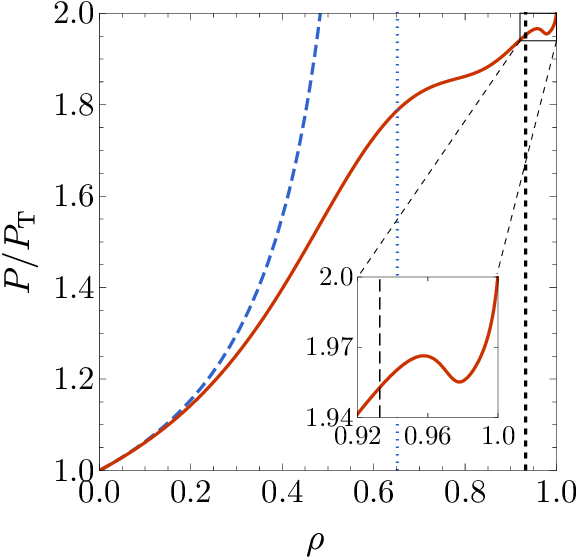}}
  \caption{
    \label{fig:EOS}
Pressure $P/P_{\rm T}$ relative to that of the Tonks gas of spherical particles with diameter $a = 1$. The solid red curve corresponds to the dumbbell system, while the dashed blue curve shows the Tonks gas with effective diameter $\langle \sigma \rangle_{\rm iso} \simeq 1.5345$. In the latter, the pressure diverges as $\rho \to 1/\langle \sigma \rangle_{\rm iso} \simeq 0.652$ (vertical dotted blue line). The vertical dashed black line indicates the density $\rho_b \simeq 0.93$ at which the ODF changes from unimodal to bimodal. The inset shows a magnified view of $P/P_{\rm T}$ in the range $0.92 < \rho < 1$.
}
\end{figure}

\subsection{Equation of state}
From Eq.~\eqref{EOS}, the density can be obtained as a function of the pressure. Nevertheless, as is customary, we present the equation of state with $\rho$ as the independent variable and $P$ as the dependent variable.

To highlight the influence of the orientational degrees of freedom, we focus on the ratio $P/P_{\rm T}$, where $P_{\rm T} = \rho / (1 - \rho)$ is the Tonks pressure for  hard rods of diameter $a = 1$, which shares the same close-packing density, $\rho_{\rm cp} = 1$, as our system.

The ratio $P/P_{\text{T}}$ is plotted  in Fig.~\ref{fig:EOS} (solid red line) as a function of density. Its deviation from unity reflects the effect of the orientational degrees of freedom and the associated  fluctuations. These deviations already appear  at low densities, where $P/P_{\text{T}}=1+(\langle\sigma\rangle_{\text{iso}}-1)\rho+\mathcal{O}(\rho^2)$. In this low-density regime, the dumbbell system is more accurately represented by a Tonks gas of  particles with diameter $\langle\sigma\rangle_{\text{iso}}$, for which $P_{\text{T,iso}}=\rho k_BT/(1-\rho \langle\sigma\rangle_{\text{iso}})$, as illustrated by the dashed blue line in Fig.~\ref{fig:EOS}.

The quantities $P/P_{\rm T}$ and $P_{\rm T,iso}/P_{\rm T}$ agree well up to $\rho \simeq 0.2$, but beyond this density they begin to separate because $P_{\rm T,iso}$ diverges at its close-packing density, $1/\langle \sigma \rangle_{\rm iso} \simeq 0.652$ (indicated by the vertical dotted blue line in Fig.~\ref{fig:EOS}). Near this density, the growth of $P/P_{\rm T}$ slows down and eventually exhibits an inflection point at $\rho \simeq 0.78$.

The change of curvature of $P/P_{\rm T}$ around $\rho = 1/\langle \sigma \rangle_{\rm iso}$ can be understood in terms of the competition between orientational and packing entropies~\cite{Mizani-Oettel-Gurin-Varga_SciPostPhysCore_2025}. 
For $\rho < 1/\langle \sigma \rangle_{\rm iso}$, orientational entropy dominates over packing entropy, whereas the reverse is true for $\rho > 1/\langle \sigma \rangle_{\rm iso}$. 
When orientational entropy prevails, the orientational distribution remains broad and only weakly structured, while the dominance of packing entropy promotes a more constrained and increasingly localized distribution around the
orientations that minimize the contact distance, in close analogy with entropy-driven orientational ordering in anisotropic-particle systems~\cite{Vroege-Lekkerkerker_RepProgPhys_1992,Luis-Kike-Yuri_JPhysCondensMatter_2014}.


As the strength of ordering increases, the average contact distance decreases from $\langle \sigma \rangle_{\rm iso} \simeq 1.5345$ toward $a = 1$, allowing the density to grow without a rapid increase of pressure, since more free space becomes available for translational fluctuations. Simultaneously, orientational fluctuations diminish, but this occurs continuously, without any singular thermodynamic signature. This contrasts with higher-dimensional systems, where the same entropy competition drives true disorder-order phase transitions. In the q1D case studied here, the only manifestation of this competition is a slight shoulder in $P/P_{\rm T}$ near $\rho = 1/\langle \sigma \rangle_{\rm iso}$.

In the high-pressure limit, as $\rho \to 1$, one finds $P/P_{\rm T} \to 2$. This shows that, although orientational fluctuations become progressively smaller as the density approaches close packing, they continue to contribute to the pressure to the same extent as positional fluctuations. This effect originates from the linear coupling between orientational and positional fluctuations near close packing~\cite{Gurin_PRE_2011}: if a dumbbell is in contact with its nearest neighbor and attempts to rotate by an angle $\delta\varphi$, it necessarily collides with that neighbor and displaces it by an amount $\delta x \propto \delta\varphi$.

This factor is not specific to the dumbbell geometry. It also appears in q1D systems of confined hard disks~\cite{Varga-Ballo-Gurin_JStatMechTheorExp_2011,MS23,V25}, where the longitudinal contact distance likewise varies linearly with small transverse deviations near close packing. More generally, whenever rotational or positional fluctuations generate positional displacements that scale linearly with the fluctuation amplitude, they contribute comparably to the pressure in the high-density limit. In related q1D systems of confined hard spheres~\cite{V25,MS25}, the same mechanism leads instead to the limiting value $P/P_{\rm T}\to 5/2$, reflecting the larger number of positional degrees of freedom involved in the collision process.

The most striking feature observed in Fig.~\ref{fig:EOS} is the nonmonotonic behavior of $P/P_{\rm T}$ around $\rho \simeq 0.96$. Note that the pressure $P$ itself remains a monotonically increasing function of density; the loop observed in Fig.~\ref{fig:EOS} simply signals a density range in which the rate of increase of the dumbbell-system pressure, relative to that of the Tonks gas, varies very rapidly. To elucidate the origin of this behavior, we next analyze the orientational ordering and the associated fluctuations.

\begin{figure}
  \resizebox{0.4\textwidth}{!}{\includegraphics{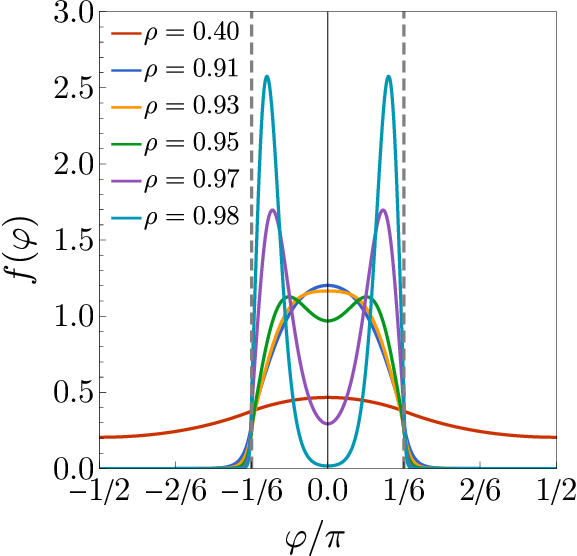}}
  \caption{Orientational distribution function (ODF), $f(\varphi)$, at densities $\rho = 0.40$, $0.91$, $0.93$, $0.95$, $0.97$, and $0.98$. The two vertical dashed lines indicate the angles $\varphi = \pm \pi/6$.
    \label{fig:ODF} }
\end{figure}

\begin{figure}
  \resizebox{0.4\textwidth}{!}{\includegraphics{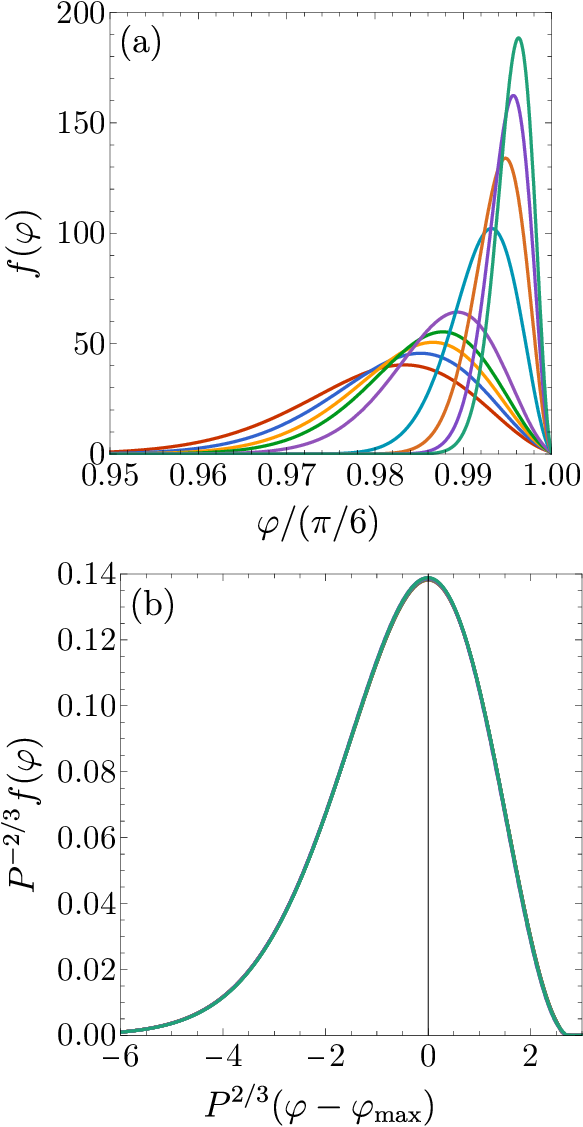}}
  \caption{(a) Orientational distribution function (ODF), $f(\varphi)$, in the vicinity of $\varphi = \pi/6$ at pressures $P = 5\times 10^3$, $6\times 10^3$, $7\times 10^3$, $8\times 10^3$,  $10^4$, $2\times 10^4$, $3\times 10^4$, $4\times 10^4$, and $5\times 10^4$.
(b) Collapse of the curves in panel (a) onto a common master curve using an appropriate scaling.
    \label{fig:ODF_scaled} }
\end{figure}

\subsection{Orientational distribution function}
\label{sec4.B}

The ODF is given by Eq.~\eqref{ODF}. It is plotted in Fig.~\ref{fig:ODF} for densities $\rho = 0.40$, $0.91$, $0.93$, $0.95$, $0.97$, and $0.98$. At low densities (represented by $\rho = 0.40$ in Fig.~\ref{fig:ODF}), the ODF is nearly uniform, $f(\varphi) \approx 1/\pi$, with a shallow maximum at $\varphi = 0$. As the density increases (see $\rho = 0.91$), the distribution becomes increasingly concentrated within the interval $-\pi/6 \le \varphi \le \pi/6$, and the maximum at $\varphi = 0$ becomes more pronounced. At $\rho = \rho_b \simeq 0.93$, this maximum flattens, and for larger densities a local minimum develops at $\varphi = 0$, flanked by two symmetric maxima. Upon further increasing the density, these maxima shift toward $\varphi = \pm \pi/6$, and $f(\varphi)$ becomes progressively more localized around these orientations.

This evolution can be understood in terms of the interplay between orientational and packing entropy. At low and moderate densities, orientational entropy dominates: fluctuations around $\varphi = 0$ are most easily accommodated because they avoid angular regions where the contact distance increases rapidly, and the ODF therefore remains unimodal. As the density increases, the system must reduce its average contact distance to sustain further compression, favoring closer alignment and signaling the growing dominance of packing entropy.

Near $\rho \simeq 0.93$, the available free volume becomes so limited that fluctuations around $\varphi = 0$ are strongly suppressed, whereas they remain comparatively less costly near the degenerate minima of the contact distance at $\varphi = \pm \pi/6$. Beyond this point, orientational entropy is maximized by concentrating probability weight near $\pm \pi/6$, where fluctuations entail a smaller loss of free volume. As a result, the ODF becomes bimodal, with two symmetric peaks that progressively shift toward $\pm \pi/6$ as $\rho \to 1$. 


The density $\rho_b \simeq 0.93$, at which $f(\varphi)$ changes from unimodal ($\rho < \rho_b$) to bimodal ($\rho > \rho_b$), is directly related to the loop observed in $P/P_{\rm T}$ in Fig.~\ref{fig:EOS}. Both features originate from the same competition between orientational and packing entropy. For densities above $\rho_b$, packing entropy can increase at the expense of orientational entropy, causing the pressure to grow more slowly with density.

We have empirically verified that, in the high-pressure regime, the ODF exhibits simple scaling behavior. Owing to the symmetry $f(\varphi)=f(-\varphi)$, we restrict the analysis to $\varphi>0$. First, the position of the maximum, $\varphi_{\max}(P)$, approaches $\varphi=\pi/6$ according to
\beq
\label{varphi_max}
\varphi_{\max}(P)\approx \frac{\pi}{6}-C_1P^{-3/5},\quad C_1\simeq 1.125.
\eeq
Next, defining the scaled angular deviation
\beq
\label{u}
u=P^{2/3}\left[\varphi-\varphi_{\max}(P)\right],
\eeq
we find that
\beq
\label{F(u)}
f(\varphi;P)\approx P^{2/3} F(u),\quad \varphi>0,
\eeq
where the scaling function $F(u)$ is independent of $P$. This data collapse is illustrated in Fig.~\ref{fig:ODF_scaled}. Note that $F(u)$ is normalized according to $\int_{-\infty}^{\infty} du\,  F(u)=1/2$. Numerically, we obtain $C_2\equiv -\langle u\rangle=-2\int_{-\infty}^{\infty} du\, u F(u)\simeq 0.5$. Consequently, in the high-pressure limit,
\beq
\langle |\varphi|\rangle \approx \frac{\pi}{6}-C_1P^{-3/5}-C_2P^{-2/3}.
\eeq

We prove in Appendix~\ref{appA} that $\lambda_0\propto e^{-P}/P^2$ in the high-pressure regime. Using $\rho^{-1}=-\partial_P\ln\lambda_0$, it is straightforward to obtain $\lim_{P\to\infty} P/P_{\text{T}}=2$, in agreement with the behavior observed in Fig.~\ref{fig:EOS}.

\begin{figure}
  \resizebox{0.4\textwidth}{!}{\includegraphics{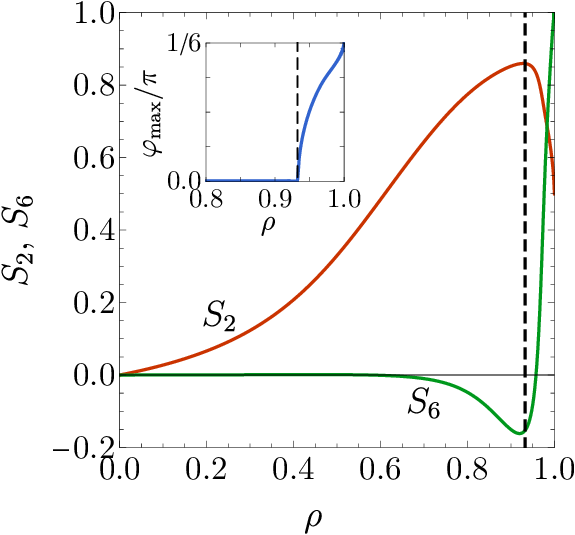}}
  \caption{Nematic order parameter $S_2$ defined by Eq.~\eqref{order_param_S2} and hexatic order parameter $S_6$ defined by Eq.~\eqref{order_param_S6} as functions of density. The inset shows the evolution of the location of the  ODF maximum: $\varphi_{\max}=0$ for $\rho<\rho_b$, $\varphi_{\max}\neq0$ for $\rho>\rho_b$, and $\varphi_{\max}\to\pm\pi/6$ as $\rho\to1$. The vertical dashed line indicates $\rho_b\simeq0.93$.
\label{fig:S}}
\end{figure}

\subsection{Order parameters}

We now examine the nature and properties of the orientational ordering that emerges in the dumbbell system, as characterized by the order parameters defined in Eqs.~\eqref{order_param_S2} and \eqref{order_param_S6}.

Both parameters are shown in Fig.~\ref{fig:S} as functions of density. In the low-density limit ($\rho\to 0$), the ODF is uniform, so that $S_2\to 0$ and $S_6\to 0$. As the density increases, orientational fluctuations around $\varphi=0$ are favored, leading to a local maximum of the ODF at this angle and a corresponding increase of $S_2$. This growth continues up to $\rho\simeq \rho_b\simeq 0.93$, where $S_2$ reaches a maximum value $S_2\simeq 0.86$. Beyond this density, tighter packing constraints become dominant and the ODF turns bimodal, with two symmetric peaks located at $\varphi_{\max}\approx\pm\pi/6$. In the close-packing limit the ODF becomes sharply localized around these two angles, so that $S_2$ approaches the limiting value $S_2\to 1/2$, characteristic of this geometry-driven q1D ordering.

The behavior of the hexatic order parameter $S_6$ is markedly different. It remains close to zero up to $\rho\simeq 0.7$, then decreases slightly to a minimum value $S_6\simeq -0.16$, and subsequently increases rapidly, reaching the limiting value $S_6=1$ at close packing. As shown in Fig.~\ref{fig:S}, the maximum of $S_2$ and the minimum of $S_6$ occur very close to the onset of the bimodal ODF at $\rho=\rho_b$.

Both the low- and high-density behavior of the ODF, and hence of $S_2$ and $S_6$, are governed by the increasing dominance of packing entropy over orientational entropy; however, the resulting ordering is qualitatively different below and above $\rho_b$. To understand the origin of this difference, it is necessary to examine what determines the location of the peak of the ODF. In the usual case where the contact distance has a unique minimum, this minimum fixes the position of the ODF peak, which coincides with the nematic director. By contrast, when the minimum contact distance is degenerate, as in the case of dumbbells and needles, the peak of the ODF cannot be determined solely by the location of the minimum. One must also consider how orientational fluctuations are maximized.

For long-wavelength orientational fluctuations, neighboring dumbbells fluctuate almost in phase. These fluctuations have a very low cost because neighboring dumbbells remain nearly parallel, so their contact distance increases only weakly. Their amplitude is limited by the requirement that the nearly common orientation of neighboring dumbbells remain within the interval $-\pi/6<\varphi\approx\varphi'<\pi/6$, beyond which the contact distance rises rapidly. As a result, long-wavelength coherent fluctuations favor $\varphi_{\max}$ as far as possible from $\pm\pi/6$, that is, $\varphi_{\max}=0$.

In contrast, the cost of short-wavelength fluctuations, where neighboring dumbbells deviate independently from the preferred direction $\varphi_{\max}$, is controlled by the local slope of the contact distance, quantified by $\tau$ (see Fig.~\ref{fig:slope_of_contact_distance}). Regions where $\tau$ is smaller allow larger orientational deviations at a smaller increase of the contact distance. Since $\tau$ decreases as $\varphi$ approaches $\pm\pi/6$, these fluctuations favor placing $\varphi_{\max}$ as close as possible to those angles, where the ODF can be broader.

Thus, the most probable orientation is not determined simply by a competition between positional fluctuations (packing entropy) and orientational fluctuations (orientational entropy), but rather by a competition between short- and long-wavelength orientational fluctuations. This competition itself is driven by the reduction of available space as the density increases. From this perspective, dumbbells differ fundamentally from needles: for needles, $\tau$ has a minimum at $\varphi=0$, so $\varphi_{\max}=0$ always maximizes orientational fluctuations and hence orientational entropy. For dumbbells, however, the usual notion of a nematic director becomes ill-defined at $\rho_b$, where two ODF peaks emerge spontaneously. As the density increases further, these peaks become narrower and move toward $\pm\pi/6$. Although the total orientational order increases, the angular separation of the peaks reduces the projection of the orientations onto the nematic director, and therefore the conventional nematic order parameter $S_2$ decreases even though the total orientational order increases.

The high-pressure behavior of $S_2$ and $S_6$ can be obtained from the scaling form of the ODF, as given by  Eqs.~\eqref{varphi_max}--\eqref{F(u)}.  Expanding the trigonometric functions around $\varphi_{\max}$ yields
\begin{subequations}
\bal
S_2\approx &2\int_{-\infty}^\infty du\, F(u)\cos\left(\frac{\pi}{3}-2C_1P^{-3/5}+2P^{-2/3}u\right)\nn
\approx&\frac{1}{2}+\sqrt{3}\left(C_1P^{-3/5}+C_2P^{-2/3}\right),
\eal
\bal
S_6\approx &-2\int_{-\infty}^\infty du\, F(u)\cos\left({\pi}-6C_1P^{-3/5}+6P^{-2/3}u\right)\nn
\approx&1-18C_1P^{-6/5}\left(C_1+2C_2P^{-1/15}\right).
\eal
\end{subequations}
Eliminating $P$ between these expressions gives the leading asymptotic relation $S_6\approx 1-6(S_2-1/2)^2$ in the close-packing limit.

\begin{figure*}
  \resizebox{0.7\textwidth}{!}{\includegraphics{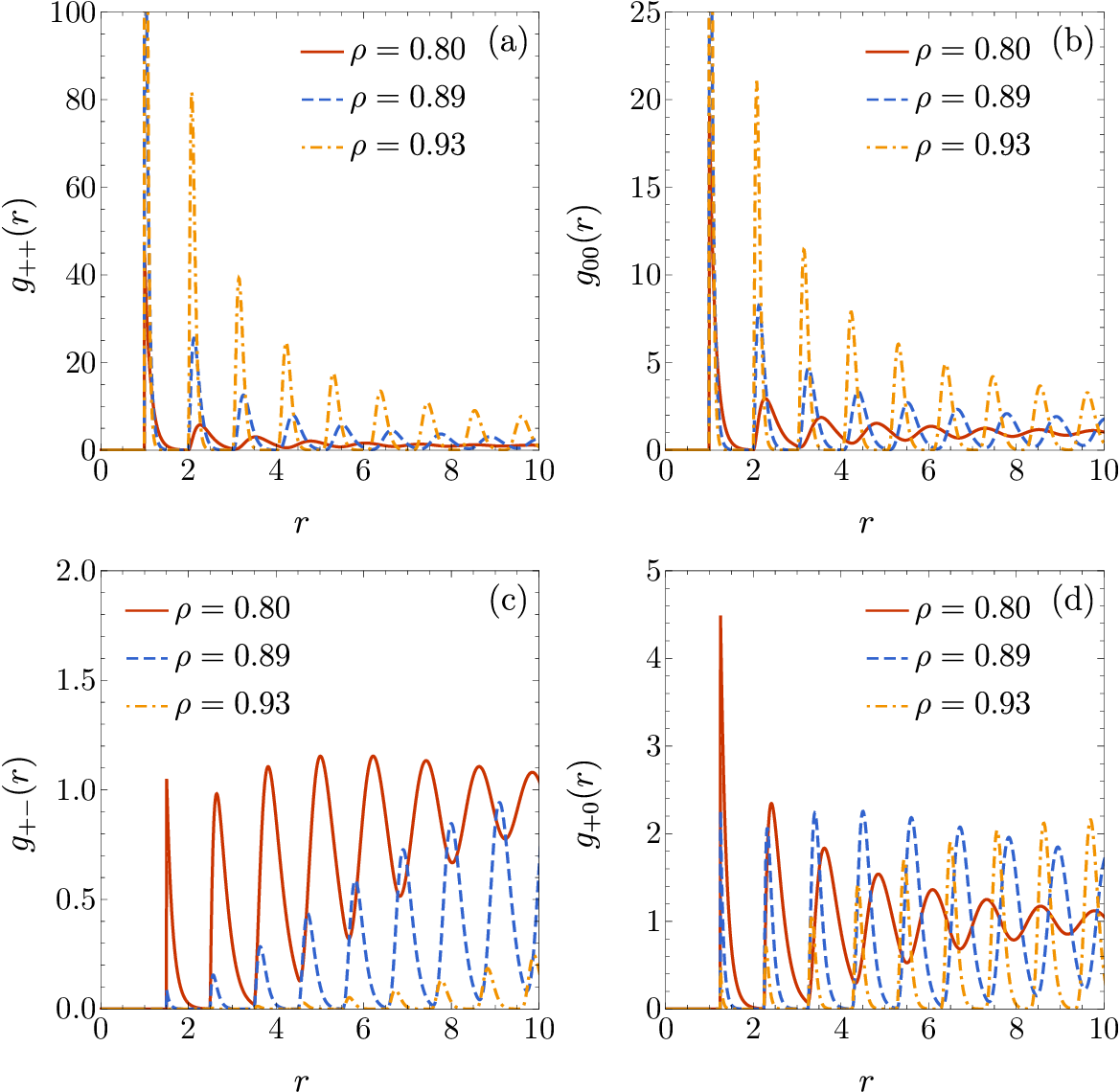}}
  \caption{Plot of (a) $g_{++}(r)$, (b) $g_{00}(r)$, (c) $g_{+-}(r)$, and (d) $g_{+0}(r)$ for $\rho=0.80$ (solid lines), $0.89$ (dashed lines), and $0.93 \simeq \rho_b$ (dash-dotted lines).
    \label{fig:gxPartial} }
\end{figure*}

\begin{figure}
 \resizebox{0.4\textwidth}{!}{\includegraphics{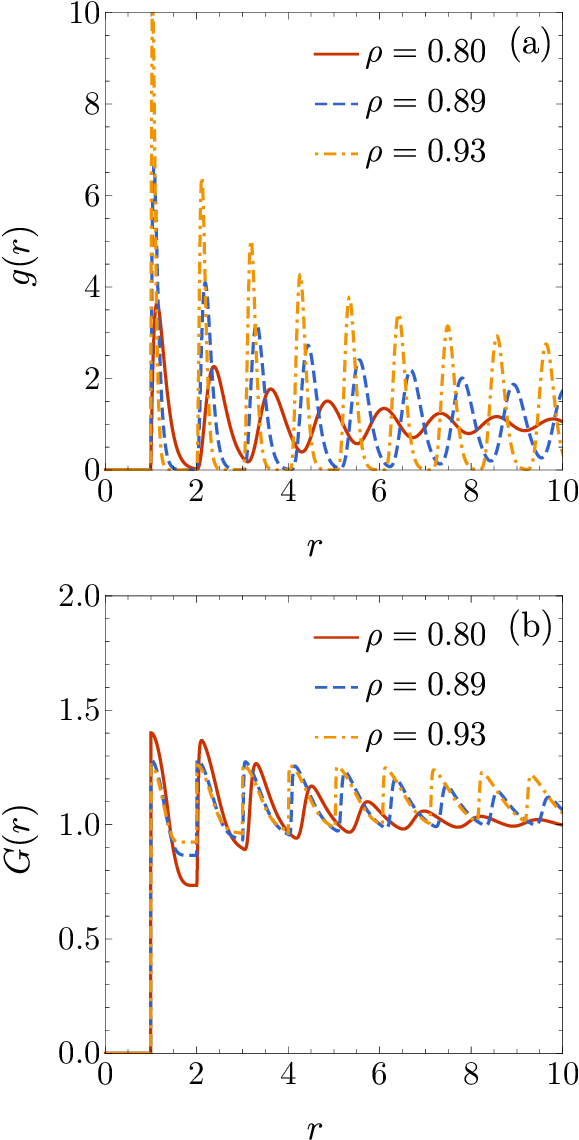}}
  \caption{
Plot of (a) the RDF $g(r)$ and (b) the distance-dependent orientational correlation function $G(r)$ for densities $\rho=0.80$ (solid lines), $0.89$ (dashed lines), and $\rho_b=0.93$ (dash-dotted lines).
    \label{fig:gx} }
\end{figure}

\subsection{Radial distribution functions}
The partial RDFs in Laplace space, $\widetilde{G}(s;\varphi,\varphi')$, are computed from the $M\times M$ discretized version of Eq.~\eqref{34}, followed by extrapolation to the continuum limit $M\to\infty$. A numerical Laplace inversion then yields $g(r;\varphi,\varphi')$.

There is, however, a simple analytic expression for the RDF in the nearest-neighbor shell, $\sigma(\varphi,\varphi')< r\leq \min_{\varphi''}[\sigma(\varphi,\varphi'')+\sigma(\varphi'',\varphi')]$, which corresponds to distances where the two dumbbells interact only with each other and no other dumbbell can fit between them. In this shell, the RDF reduces to
\beq
\label{50}
g(r;\varphi,\varphi')=\frac{e^{-P r}}{\rho\lambda_0\psi_0(\varphi)\psi_0(\varphi')}.
\eeq
The contact values are then
\beq
\label{51}
g^\cont(\varphi,\varphi')=\frac{e^{-P\sigma(\varphi,\varphi')}}{\rho\lambda_0 \sqrt{f(\varphi)f(\varphi')}}.
\eeq
Physically, $g^\cont(\varphi,\varphi')$ quantifies the ``rigidity'' of a nearest-neighbor configuration $(\varphi,\varphi')$: larger values indicate that the pair is more likely to be found at contact and less free space is available for fluctuations, while smaller values indicate more flexibility in relative orientations and separations.

As characteristic orientations, we choose $\varphi,\varphi'=0,\pm{\pi}/{6}$. Accordingly, we define
\begin{subequations}
\label{51++}
\beq
g_{++}(r)\equiv g\left(r;\frac{\pi}{6}, \frac{\pi}{6}\right)=g\left(r;-\frac{\pi}{6}, -\frac{\pi}{6}\right),
\eeq
\beq
g_{00}(r)\equiv g\left(r;0, 0\right),
\eeq
\beq
g_{+-}(r)\equiv g\left(r;\frac{\pi}{6}, -\frac{\pi}{6}\right)=g\left(r;-\frac{\pi}{6}, \frac{\pi}{6}\right),
\eeq
\beq
g_{+0}(r)\equiv g\left(r;\frac{\pi}{6}, 0\right)=g\left(r;-\frac{\pi}{6}, 0\right).
\eeq
\end{subequations}
The corresponding contact distances are  $\sigma(\varphi,\varphi')=1$ for $g_{++}$ and $g_{00}$, while $\sigma(\varphi,\varphi')=3/2$ for $g_{+-}$ and $\sigma(\varphi,\varphi')=\sigma_{+0}\equiv (1+\sqrt{9+4\sqrt{3}})/4 \simeq 1.25$ for $g_{+0}$.

The four functions defined in Eqs.~\eqref{51++} are plotted in Fig.~\ref{fig:gxPartial} for $\rho=0.80$, $0.89$, and $0.93 \simeq \rho_b$. At the first two densities, the ODF is unimodal, while for $\rho>\rho_b$, it becomes bimodal. Thus, Eq.~\eqref{51} shows that along the degenerate line $-{\pi}/{6}\leq \varphi=\varphi'\leq {\pi}/{6}$, the contact value $g^\cont(\varphi,\varphi)$ increases monotonically from $\varphi=0$ to $\varphi={\pi}/{6}$ for $\rho\leq \rho_b$. This trend is reflected in the other peaks of the RDF, as seen in Figs.~\ref{fig:gxPartial}(a) and \ref{fig:gxPartial}(b).

For densities above $\rho_b$, the rigidity parameter $g^\cont(\varphi,\varphi)$ develops a nonmonotonic profile, with a local maximum at $\varphi=0$ and symmetric minima at $\varphi=\pm\varphi_{\max}$. This can be understood by considering the competition between neighbor alignment and local flexibility. At high density, a dumbbell oriented at $\varphi=0$ is likely to have a neighbor also at $\varphi'=0$, because deviations are strongly penalized in terms of the contact distance $\sigma(\varphi,\varphi')$, creating a maximum in the pair rigidity. If the reference dumbbell rotates slightly, $\varphi\neq 0$, its neighbor tends to follow, $\varphi'=\varphi$, but the rigidity decreases. Approaching $\varphi={\pi}/{6}$, further fluctuations of the neighbor are strongly suppressed, causing the rigidity to rise again. These competing effects produce minima at $\varphi=\varphi'=\pm \varphi_{\max}$. Because $g^\cont(\varphi,\varphi)\propto 1/f(\varphi)$, this argument also explains the bimodal shape of the ODF.

Figures \ref{fig:gxPartial}(c) and (d) show that, for a reference dumbbell at $\varphi={\pi}/{6}$, the probability of finding a nearest neighbor at $\varphi'=-{\pi}/{6}$ or $\varphi'=0$ decreases sharply with increasing density, especially for $\varphi'=-{\pi}/{6}$. At larger separations, however, both $g_{+-}(r)$ and $g_{00}(r)$ converge to unity, reflecting the loss of orientational correlation at long distances.

The behavior of the partial RDFs discussed above directly influences the total RDF, $g(r)$, and the distance-dependent orientational correlation function, $G(r)$, shown in Fig.~\ref{fig:gx} for the same densities as in Fig.~\ref{fig:gxPartial}. As density increases, the peaks of both $g(r)$ and $G(r)$ shift toward integer values of $r$ and become more pronounced. This reflects the growing dominance of quasiparallel configurations, where neighboring dumbbells preferentially adopt orientations $\varphi\lesssim {\pi}/{6}$ or $\varphi\gtrsim -{\pi}/{6}$. In other words, the local ordering captured by the partial RDFs propagates through the system, giving rise to increasingly correlated positional and orientational structures at higher densities.

\begin{figure}
  \resizebox{0.4\textwidth}{!}{\includegraphics{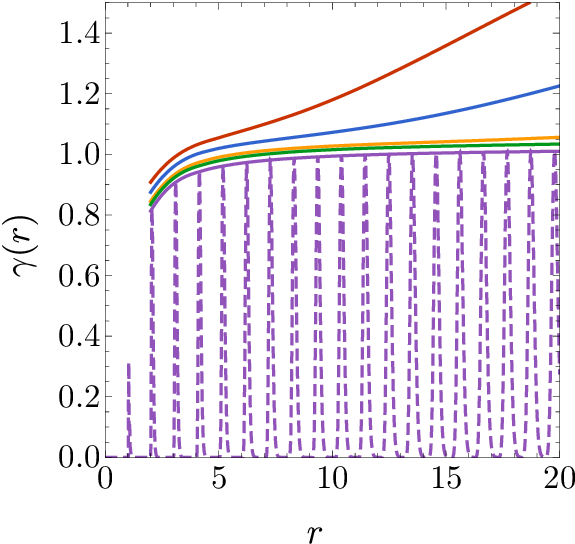}}
  \caption{Plot of the rescaled RDF $\gamma(r)$ [see Eq.~\eqref{gamma}] for $\rho=0.96$ (dashed line).
The envelopes of $\gamma(r)$ for $\rho=0.80$, $0.85$, $0.90$, $0.92$, and $0.96$ (solid lines, from top to bottom) are also shown.
\label{fig:algebraic_decay} }
\end{figure}

According to Eq.~\eqref{41}, the asymptotic decay of $g(r)-1$ is exponential (either monotonic or oscillatory), $g(r)-1\sim e^{s_1^{(0)}r}$ with $\mathrm{Re}[s_1^{(0)}]<0$.
However, in analogy with the cases of the Tonks gas~\cite{BM25}, 1D binary mixtures~\cite{MS26}, and q1D confined hard disks~\cite{HBPT20,HC21,MS26}, one may anticipate an extended intermediate regime where the decay is well described by a power law $g(r)\sim 1/\sqrt{r-1}$.
In particular, by extending the results obtained for confined hard disks~\cite{MS26}, one can expect that, for sufficiently high densities, the envelope of the rescaled RDF
\beq
\label{gamma}
\gamma(r)\equiv(1-\rho)\sqrt{\pi(r-1)}g(r)
\eeq
remains close to unity over an intermediate range of distances.

This behavior is illustrated in Fig.~\ref{fig:algebraic_decay}, which shows $\gamma(r)$ for $\rho=0.96$, together with the envelopes of $\gamma(r)$ for $\rho=0.80$, $0.85$, $0.90$, $0.92$, and $0.96$. The densities $\rho=0.80$ and $0.85$ are not large enough to exhibit an intermediate algebraic regime. For the other higher densities, however, the envelope of $\gamma(r)$ is nearly constant in the region shown, with its value approaching unity as the density increases.
This algebraic regime is a pre-asymptotic crossover that eventually gives way to the exponential decay predicted by the pole analysis.

\begin{figure}
  \resizebox{0.4\textwidth}{!}{\includegraphics{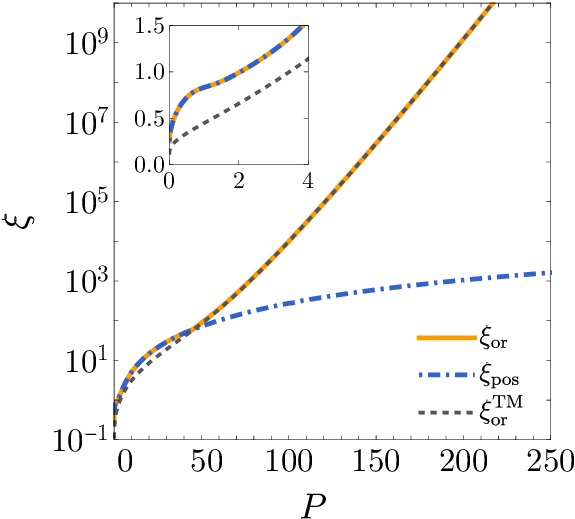}}
  \caption{Orientational correlation lengths $\xior^\tm$ (dashed line) and $\xior$ (solid line), and positional correlation length $\xipos$ (dash-dotted line) as functions of pressure.  Note that $\xior$ exhibits a kink at $P\simeq 44.7$. The inset shows a magnified view of the curves for $P\leq 4$.
      \label{fig:corr_lengths} }
\end{figure}

\subsection{Correlation lengths}
\label{sec4E}

The orientational correlation length can be defined in two complementary ways. From the TM method, $\xior^\tm$ [Eq.~\eqref{25}] measures the characteristic number of intermediate dumbbells required for a function $A(\varphi)$ to decorrelate between two separated particles, so that $G_A(n)\approx 0$ for $n\gg \xior^\tm$. Alternatively, one can define a distance-based orientational correlation length, $\xior$, characterizing the decay of correlations in space: $g(r;\varphi,\varphi')\approx 1$ and $G(r)\approx 0$ if $r \gg \xior$.

From the pole representation in Eq.~\eqref{38}, the asymptotic behavior of $g(r;\varphi,\varphi')$ is governed by the pole with the least negative real part. For $\rho<\rho_\xi\simeq 0.96$ (or, equivalently, $P<P_\xi\simeq 44.7$), this pole is the complex conjugate pair $s_1^{(0)}=-\kappa_1^{(0)}\pm\imath\omega_1^{(0)}$, implying an oscillatory decay of both $g(r;\varphi,\varphi')$ and $G(r)$. For $\rho>\rho_\xi$, the dominant contribution instead comes from the real pole $s_1^{(1)}=-\kappa_1^{(1)}$, leading to a monotonic decay. Accordingly,
\beq
\xior=\begin{cases}
 \displaystyle{\frac{1}{\kappa_1^{(0)}}},&\rho< \rho_\xi,\\
\displaystyle{\frac{1}{\kappa_1^{(1)}}},&\rho> \rho_\xi.
\end{cases}
\eeq
Thus, $\xior$ exhibits a kink at the crossover density $\rho_\xi$. It is interesting to note that $\rho_\xi$ lies only slightly above $\rho_b\simeq 0.93$.

The characteristic distance over which the total RDF, $g(r)$, decays defines the positional correlation length, $\xipos$. Since only the even poles contribute to $g(r)$ [see Eq.~\eqref{41}], the dominant contribution arises from the complex conjugate pair $s_1^{(0)}=-\kappa_1^{(0)}\pm\imath\omega_1^{(0)}$. Its real part determines the positional correlation length,
\beq
\xipos=\frac{1}{\kappa_1^{(0)}},
\eeq
while the imaginary part sets the wavelength of the oscillatory decay. Consequently, $\xipos=\xior$ for $\rho<\rho_\xi$, whereas $\xipos<\xior$ for $\rho>\rho_\xi$, reflecting the fact that, above the crossover density $\rho_\xi$, orientational correlations persist over longer distances than positional correlations.

Based on the simplified three-state model discussed in Appendix~\ref{appB}, we conjecture that, in the high-pressure limit, the leading eigenvalues $\lambda_0(s+P)$ and $\lambda_1(s+P)$ take the form
\beq
\lambda_k(s+P)\propto \frac{e^{-(s+P)}}{(s+P)^2}\left[1+\Delta_k(s+P)\right],
\eeq
with
\begin{subequations}
\beq
\Delta_k(s+P)\ll 1,
\eeq
\beq
\Delta_0(s+P)-\Delta_1(s+P)\sim e^{-\mu(s+P)}.
\eeq
\end{subequations}
The TM eigenvalues correspond to $s=0$, so that
\beq
\frac{\lambda_0}{\lambda_1}\approx 1+\Delta_0(P)-\Delta_1(P).
\eeq
Using Eq.~\eqref{25}, we then obtain
\beq
\label{xiorTM}
\xior^\tm\approx\frac{1}{\Delta_0(P)-\Delta_1(P)}
\sim e^{\mu P}.
\eeq

We now turn to the evaluation of $\xior$. According to Eq.~\eqref{43.0},
\beq
\lambda_1(-\xior^{-1}+P)=\lambda_0,
\eeq
which can be rewritten as
\beq
\label{B.8}
e^{1/\xior}\left[1+\Delta_1(P-\xior^{-1})\right]=\left(1-\frac{\xior^{-1}}{P}\right)^2\left[1+\Delta_0(P)\right].
\eeq
On physical grounds, $\xior^{-1}\to 0$ in the high-pressure limit. Expanding both sides of Eq.~\eqref{B.8} to first order in $\xior^{-1}$ yields
\bal
\label{B9}
\xior\approx&\frac{1+\Delta_1(P)-\Delta_1'(P)+2\frac{1+\Delta_0(P)}{P}}{\Delta_0(P)-\Delta_1(P)}\nn
\approx&\frac{1}{\Delta_0(P)-\Delta_1(P)}\approx\xior^\tm.
\eal
Thus, $\xior$ and $\xior^\tm$ are asymptotically equivalent in the high-pressure regime.

We next consider the positional correlation length $\xipos$, which is determined by the real part of the leading solution of
\beq
\lambda_0(s+P)=\lambda_0,
\eeq
or, equivalently,
\beq
\label{B.12}
e^{-s}\left[1+\Delta_0(s+P)\right]=\left(1+\frac{s}{P}\right)^2\left[1+\Delta_0(P)\right].
\eeq
In the limit $P\to\infty$, the correction terms $\Delta_0(P)$ and $\Delta_0(s+P)$ can be neglected with respect to unity, reducing Eq.~\eqref{B.12} to
\beq
\label{B.13}
e^{-s}=1+2\frac{s}{P}.
\eeq
This is the same equation governing the poles of a system of hard rods of unit length at pressure $P/2$. The leading solution is $s_1^{(0)}=-\xipos^{-1}\pm\imath\omega_1^{(0)}$, with
\beq
\label{xipos}
\xipos\approx\frac{P^2}{8\pi^2},\quad \omega_1^{(0)}\approx 2\pi\left(1-\frac{2}{P}\right).
\eeq

We recall that $\xior^\tm$ measures the range of orientational correlations in terms of the number $n$ of dumbbells, whereas $\xior$ measures the same range in terms of the spatial distance $r$. Their equivalence reflects the fact that, at high pressure, dumbbells predominantly adopt the parallel orientations $\varphi\approx\pm{\pi}/{6}$, so that the distance to the $n$th neighbor scales as $r\simeq n a$.

These properties are confirmed by Fig.~\ref{fig:corr_lengths}. In particular, the exponential growth $\xior^\tm \approx \xior \sim e^{\mu P}$ predicted in Eq.~\eqref{xiorTM} is clearly observed numerically, with $\mu\simeq 0.12$. Similarly, the increase of $\xipos$ is accurately captured by the quadratic scaling given in Eq.~\eqref{xipos}. The inset of Fig.~\ref{fig:corr_lengths} further shows that $\xior>\xior^\tm$ in the low-pressure regime.

This behavior also clarifies the physical meaning of the bimodal ODF discussed in Sec.~\ref{sec4.B}. A bimodal distribution could in principle arise in two different ways: either each system selects one of the two orientations $\varphi\approx\pm\pi/6$, or the system contains regions with the two orientations simultaneously. The correlation functions show that the second scenario applies here. At high densities the system is essentially composed of domains of dumbbells with nearly identical orientation (either $\varphi\approx\pi/6$ or $\varphi\approx-\pi/6$), and $\xior$ characterizes the typical spatial extent of such domains. This interpretation is further supported by the observation that the minimum distance required for $g_{+-}(r)$ to approach unity is of the order of $\xior$.

By contrast, $\xipos$ measures positional correlations within a given domain. Its much weaker growth reflects the fact that positional correlations develop only locally inside regions of nearly uniform orientation, rather than over the much larger distances associated with orientational correlations.

\section{Conclusions}
\label{sec5}

We have presented an exact analysis of a q1D fluid of hard dumbbells with a continuous orientational degree of freedom. By exploiting the bisymmetric structure of the interaction kernel, we derived the equation of state, the ODF, and both positional and orientational correlation functions. The formalism applies generally to other q1D systems with continuous internal variables and hard-core constraints.

The equation of state reveals a nontrivial interplay between orientational and packing entropy. While the pressure remains a monotonic function of density, the ratio $P/P_{\text{T}}$ exhibits a pronounced shoulder and a narrow loop-like feature at high density. These features are traced to a continuous reorganization of the orientational degrees of freedom rather than to any thermodynamic singularity. In the close-packing limit, orientational and positional fluctuations contribute equally to the pressure, leading to the asymptotic result $P/P_{\text{T}}\to 2$.

The ODF undergoes a continuous yet abrupt crossover from a unimodal shape centered at $\varphi=0$ to a bimodal form with maxima approaching $\varphi=\pm\pi/6$. The density $\rho_b\simeq 0.93$ at which this crossover takes place coincides with the onset of the loop in $P/P_{\text{T}}$ and with a maximum of the nematic-like order parameter $S_2$ and a minimum of the hexatic-like order parameter $S_6$. This behavior reflects a purely geometric mechanism: at high densities, orientational fluctuations around $\varphi=0$ are suppressed, while fluctuations near the degenerate minima of the contact distance become entropically favored. The resulting ordering corresponds to a continuous reorganization of the orientational distribution from a single preferred orientation to two symmetric ones. 
Although this change bears some analogy to a nematic--tetratic transition, it occurs here as a smooth structural crossover, consistent with the absence of true phase transitions in one dimension.

The analysis of partial and total RDFs shows how this orientational reorganization is encoded in real-space correlations. Contact values provide a natural measure of pair rigidity and directly mirror the structure of the ODF. Above $\rho_b$, the rigidity along the degenerate orientational line becomes nonmonotonic, with minima located at the most probable orientations. This mechanism explains the emergence of bimodality in the ODF and its connection to the structural features observed in the RDFs. In addition, although the asymptotic decay of the RDF is purely exponential, as dictated by the pole structure of the transfer-matrix formulation, we have identified an extended intermediate regime at high densities where the correlations follow an approximate algebraic envelope before crossing over to the asymptotic exponential behavior. This feature is consistent with similar intermediate regimes reported in other q1D hard-core systems.

We have also characterized correlation lengths associated with orientational and positional order. The orientational correlation length exhibits a kink separating oscillatory and monotonic decay regimes, reflecting a change in the nature of the leading pole of the Laplace-transformed correlation functions. In contrast, the positional correlation length grows algebraically at high pressure. In this regime, the orientational correlation length grows exponentially with pressure and coincides with the TM correlation length, indicating the formation of extended domains of nearly parallel dumbbells. A simplified three-state model captures these behaviors and clarifies their physical origin.

Overall, this work highlights how purely entropic effects, combined with geometric constraints, can produce rich structural and correlation phenomena in q1D systems. The exact results obtained here provide a useful benchmark for approximate theories and simulations, and suggest that similar mechanisms may operate in other confined fluids with internal degrees of freedom. The transfer-matrix formulation developed here could also be extended to include external fields acting on the translational or orientational degrees of freedom through additional one-body contributions to the transfer kernel.

\begin{acknowledgments}
A.M.M. and A.S acknowledge financial support from Grant No.~PID2024-156352NB-I00 funded by MCIU/AEI/10.13039/501100011033 and by ERDF/EU, and from Grant No.~GR24022 funded by the Junta de Extremadura (Spain). A.M.M. is grateful to the Spanish Ministerio de Ciencia e Innovaci\'on for a fellowship PRE2021-097702.
P.G. and S.V. gratefully acknowledge the financial support of the National Research, Development and Innovation Office---Grants No.\ 2023-1.2.4-T\'ET-2023-00007, No.\ NKFIH K137720, and No.\ TKP2021-NKTA-21.
\end{acknowledgments}

\section*{Data availability}
The data supporting the findings of this study are available from the corresponding author upon reasonable request. In addition, the open-source C++ code used to obtain the results presented in this paper is available through Ref.~\cite{Dumbbells_github}.

\appendix
\section{High-pressure limit of $\lambda_0$}
\label{appA}
From Eq.~\eqref{12} we have
\bal
\label{A1}
\lambda_0=&\int_{-\frac{\pi}{2}}^{\frac{\pi}{2}} d\varphip
     \int_{2|\varphip|-\pi}^{\pi-2|\varphip|} d\varphim \,\psi_0\left(\varphip-\frac{\varphim}{2}\right)   \psi_0\left(\varphip+\frac{\varphim}{2}\right)        \nn
   & \times
          \frac{e^{-P \sigma\left(\varphip-\frac{\varphim}{2},\varphip+\frac{\varphim}{2}\right)}}{P},
\eal
where
\beq
\varphip\equiv  \frac{\varphi + \varphi'}{2},\quad
\varphim\equiv  \varphi' - \varphi.
 \label{change_of_variables}
\eeq

In the limit $P\to\infty$, the product $\psi_0(\varphi)\psi_0(\varphi')$ becomes strongly localized near $\varphi=\varphi'=\pm{\pi}/{6}$ (see Fig.~\ref{fig:ODF_scaled}). As a consequence, $|\varphim|$ can be treated as a small parameter and $\varphip\approx\pm{\pi}/{6}$. This allows us to make the following approximations:
\begin{subequations}
\beq
\int_{-\frac{\pi}{2}}^{\frac{\pi}{2}} d\varphip\int_{2|\varphip|-\pi}^{\pi-2|\varphip|} d\varphim\,\cdots\approx \int_{-\infty}^{\infty} d\varphip\int_{-\infty}^{\infty} d\varphim\,\cdots,
\eeq
\beq
\sigma\left(\varphip-\frac{\varphim}{2},\varphip+\frac{\varphim}{2}\right)\approx 1+c|\varphim|,\quad c=\frac{\sqrt{3}}{4},
\eeq
\beq
\psi_0\left(\varphip-\frac{\varphim}{2}\right)\psi_0\left(\varphip+\frac{\varphim}{2}\right)\approx f(\varphip)\left[1+\frac{\varphim^2}{8}\frac{\partial^2\ln f(\varphip)}{\partial\varphip^2}\right].
\eeq
\end{subequations}
Inserting these approximations into Eq.~\eqref{A1} yields
\bal
\label{A4}
\lambda_0\approx&\frac{e^{-P}}{P}\int_{-\infty}^{\infty} d\varphip f(\varphip)
     \int_{-\infty}^{\infty} d\varphim \,e^{-P c|\varphim|}
     \nn   & \times
          \left[1+\frac{\varphim^2}{8}\frac{\partial^2\ln f(\varphip)}{\partial\varphip^2}\right]\nn
                        =&
          \frac{2e^{-P}}{P^2c}\int_{-\infty}^{\infty} d\varphip f(\varphip)
               \left[1+\frac{1}{4(P c)^2}\frac{\partial^2\ln f(\varphip)}{\partial\varphip^2}\right]\nn
=&\frac{2e^{-P}}{P^2c}\left\{1-\frac{1}{4(P c)^2}\int_{-\infty}^{\infty} d\varphip\, f(\varphip)\left[\frac{\partial\ln f(\varphip)}{\partial\varphip}\right]^2\right\},
\eal
where, in the last step, we have integrated by parts.

Using the scaling relations in Eqs.~\eqref{u} and \eqref{F(u)}, we estimate
\beq
\int_{-\infty}^{\infty} d\varphip\, f(\varphip)\left[\frac{\partial\ln f(\varphip)}{\partial\varphip}\right]^2\approx P^{4/3}\int_{-\infty}^{\infty} du\, \frac{\left[F'(u)\right]^2}{F(u)}.
\eeq
Since this contribution is subleading with respect to the prefactor $P^{-2}$, we finally obtain
\beq
\lambda_0\approx\frac{2e^{-P}}{P^2c}.
\eeq

\section{Correlation lengths in a  three-state model}
\label{appB}
To illustrate the main properties of the correlation lengths $\xior^\tm$, $\xior$, and $\xipos$, we consider a toy three-state model in which the only allowed orientations are $\varphi=-{\pi}/{6}$, $0$, and ${\pi}/{6}$. The corresponding contact-distance matrix reads
\beq
\sigma(\varphi,\varphi')
=\begin{pmatrix}
  1&b&\frac{3}{2}\\
  b&1&b\\
  \frac{3}{2}&b&1
\end{pmatrix},
\eeq
where $b=\sigma_{+0}=(1 + \sqrt{9 + 4 \sqrt{3}})/{4}\simeq 1.25$.

The eigenvalues of the matrix $\widehat{\Omega}(s)$ are
\beq
\lambda_k(s)=\frac{e^{-s}}{s}\left[1+\Delta_k(s)\right],
\eeq
with
\begin{subequations}
\label{13}
\beq
\Delta_0(s)=\frac{e^{-s/2}}{2}+\sqrt{2e^{-2(b-1)s}+\frac{e^{-s}}{4}},
\eeq
\beq
\Delta_1(s)=-e^{-s/2},
\eeq
\beq
\Delta_2(s)=\frac{e^{-s/2}}{2}-\sqrt{2e^{-2(b-1)s}+\frac{e^{-s}}{4}}.
\eeq
\end{subequations}

In the high-pressure limit, these expressions simplify to
\begin{subequations}
\beq
\Delta_0(s+P)\approx\sqrt{2}e^{-(b-1)(s+P)},
\eeq
\beq
\Delta_1(s+P)\approx-e^{-(s+P)/2},
\eeq
\beq
\Delta_2(s+P)\approx-\sqrt{2}e^{-(b-1)(s+P)}.
\eeq
\end{subequations}

Following steps analogous to those in Sec.~\ref{sec3}, one obtains
\begin{subequations}
\beq
\xior^\tm\approx\xior\approx\frac{1}{\sqrt{2}}e^{(b-1)P}\left[1-\frac{e^{-(3/2-b)P}}{\sqrt{2}}\right],
\eeq
\beq
\xipos\approx\frac{P^2}{2\pi^2},\quad \omega_1^{(0)}\approx 2\pi\left(1-\frac{1}{P}\right).
\eeq
\end{subequations}

Apart from these asymptotic results, it is worth noting that in the three-state model the kink in $\xior$ appears at a much lower pressure than in the continuous case, namely at $P_\xi\simeq 2.66$, corresponding to a density $\rho_\xi\simeq 0.66$.

%
\end{document}